\documentclass[preprint,secnumarabic,amssymb, nobibnotes, aip, pof]{revtex4-1}
\usepackage{graphicx}
\usepackage{amsmath}

\usepackage{graphicx}
\usepackage{dcolumn}
\usepackage{bm}
\usepackage[usenames, dvipsnames]{color}
\usepackage[table,xcdraw]{xcolor}
\definecolor{pink}{rgb}{0.858, 0.188, 0.478}

\definecolor{green}{RGB}{205,100,60}

\begin{document}

\title{Small scale dynamics of a shearless turbulent/non-turbulent interface in dilute polymer solutions}%

\begin{abstract}

We study the physics of turbulent/non-turbulent interface (TNTI) of an isolated turbulent region in dilute polymer solutions and Newtonian fluid. 
We designed an experimental setup of a turbulent patch growing in water/dilute polymer solution, without mean shear and far from the walls. The observations from the experiments are complemented and expanded by simulations performed using a localised homogeneous forcing to generate the turbulent front and the FENE-P model for the polymer stress. The comparison, 
which shows that when Newtonian and viscoelastic TNTIs are fed by the same energy they behave in similar manner both in the experiments and in the simulations, permits to extend the applicability, on a qualitative basis, of single relaxation time polymer models also to turbulent/non-turbulent interfaces. From the detailed analysis offered by the numerical results, the alterations in the dynamics between strain and vorticity help understanding the mechanics of the polymer action on the TNTI without mean shear. The reduced vorticity stretching and increased vorticity compression terms are found to be due to the  modified degrees of alignment between vorticity, polymer conformation tensor and  rate-of-strain tensor eigenvectors observed especially near the interface. These alignments at the smallest scales of the non-Newtonian turbulent flow lead to a reduced production of enstrophy and consequently to a reduced entrainment, that in this problem are seen as reduced advancement of a turbulent region.

\end{abstract}
\author{G. Cocconi}
 \affiliation{Institute of Fluid Mechanics, Karlsruhe Institute of Technology, 76131, Germany}

\author{E. De Angelis}%
\affiliation{School of Engineering - Cardiff University, Queen's Buildings, The Parade,Cardiff CF24 3AA, United Kingdom
}%

\author{B. Frohnapfel}
 \affiliation{Institute of Fluid Mechanics, Karlsruhe Institute of Technology, 76131, Germany}

\author{M. Baevsky}%
\affiliation{ School of Mechanical Engineering,Tel-Aviv University,Tel-Aviv 69978, Israel}
 
\author{A. Liberzon}%
 \email{alexlib@tau.ac.il}
\affiliation{ School of Mechanical Engineering,Tel-Aviv University,Tel-Aviv 69978, Israel}%

\maketitle

\date{\today}
\maketitle

\preprint{POF/16-1470-LB}

\section{Introduction}

Turbulent/non-turbulent interfaces (TNTIs) are sharp boundaries between regions of rotational and irrotational fluctuations of velocity. These interfacial layers play a central role in the mixing of scalar (temperature, concentration) and dynamical (momentum and vorticity) quantities and are omnipresent in turbulent jets, wakes and boundary layers. 
The study of turbulent interfaces is challenging for theoretical, numerical or experimental methods~\cite{da_silva_2014rev}, since they are unsteady and strongly convoluted surfaces with fractal features spanning over a broad range of scales.  At the interface, large and small scales dynamics contribute to the entrainment of non-turbulent fluid into the mass of the turbulent flow. Since Corrsin and Kistler~\cite{corrsin_1955} it is recognized that fluctuating vorticity can propagate only via small scale viscous diffusion which is the driving mechanism of propagation in a thin layer at the very edge of the TNTI. Such viscous layer, with thickness of the order of the Kolmogorov scale, was observed to be a bounding edge of a thicker layer of the order of the Taylor microscale. Within this turbulent layer vorticity decays from the levels found in the bulk of the flow to the very small values of the viscous layer~\cite{da_silva_2014rev}.  Large scales in the turbulent bulk affect the entrainment across the interface by the flux of momentum and vorticity and via strong distortion and convolution of the thin viscous layer. Large scale mean shear also appears to enhance entrainment by increasing both the viscous and the inviscid contributions~\cite{wolf2013}.

A study of the effects of dilute polymers on a TNTI can be an important step toward better understanding of both the polymer dynamics in turbulence and of the small scale dynamics of the interfaces. Dilute polymer solutions are known to produce macroscopic changes in turbulent flows through interactions between the smallest velocity gradients and the polymer chains~\cite{DeAngelis2005,crawford2008,jovanovic2005,Dubief_2004}. Such effects in homogeneous isotropic turbulence were inferred to relate to the polymers orientation with respect to the small scale velocity gradients~\cite{Liberzon2005,Davoudi06}.  
Simulations with  polymer models have shown a strong tendency of polymer conformation tensor eigenvectors to align with the fluctuating vorticity vector and with the stretching eigenvector of the rate-of-strain tensor~\cite{Vincenzi2015,Valente_2014}. It has been also observed that polymers change the flow through interactions with coherent structures\cite{Dubief_2004,bagheri2012,perlekar2010}. Since TNTI is a well defined region of 3D turbulent flow with a thicker layer of coherent motions bounded by a strongly viscous layer~\cite{da_silva_2014rev}, it is an ideal flow state to study the dynamics of polymers and their interaction with the turbulent fluctuations. In the past, it was shown that the entrainment in a flow without changing its energy injection mechanism is altered by the addition of dilute polymers~\cite{Liberzon2009}. However, this experimental study was limited to the large scales and the results could have been contaminated by the presence of strong shear layers at the side walls of the oscillating grid tank. 

We propose here to extend that study to explore the interaction of polymers with the small scale features of the flow and the effects on the turbulent dynamics of the TNTI. Despite several attempts to directly visualize the dynamics of polymer molecules in simple shearing flows~\cite{graham2011}, up to today it is not possible to measure the extension and orientation of polymers in turbulent flow experiments. Therefore, we combine the experimental study with direct numerical simulations (DNS) that can reveal the underlying dynamics of the polymers. DNS of turbulent flows with polymers has commonly to rely on simplified single relaxation time polymer models that are known to replicate at best only qualitative aspects of actual turbulent flows with dilute polymer solutions \cite{Vincenzi2015,ghosh2001}. Hence their application can give useful physical insight only for the cases where experimental evidence confirms the observed trends in simulations. A joined experimental and numerical approach can give detailed information on the dynamics of polymers near the TNTI while reducing the uncertainties related to the utilisation of polymer models in simulations.  

Based on our previous experience with dilute polymer solutions~\cite{DeAngelis2005, Casciola2007a, Liberzon2005, Liberzon2009}, we propose to address the problem of TNTI by analyzing the growth of a localized turbulent region into a non-turbulent fluid in homogeneous solutions of dilute polymers and a Newtonian fluid. There are several possible configurations in which a comparative study can be performed. For instance we can compare two transient cases that propagate at different rates~\cite{Liberzon2009}. In such case the kinetic energy in the turbulent patch is a free parameter and it hinders our ability to compare the small scale dynamics near the interface. In this work we address the comparison differently - we first run an extensive set of experimental runs in Newtonian and dilute polymer solutions of different concentration. Then we select the flow cases in which the turbulent kinetic energy within the turbulent region is comparable. In the numerical part we run a Newtonian flow case first and then tune the energy input in the polymer case in order to obtain a comparable data set. Such configuration provides a unique view into the dynamics of turbulent flows with and without dilute polymers near the interface, under equivalent turbulent kinetic energy conditions.

The paper is organized as follows. We describe in details our experimental setup and numerical simulations in Section~\ref{sec:methods}. We present the key results in Section~\ref{sec:results}, followed by the conclusions in Section~\ref{sec:conclusions}.

\section{Methods}\label{sec:methods} 

We study the small scale dynamics of the dilute polymer solution near the TNTI using DNS with the Finitely Extensible Elastic model with the Peterlin closure (FENE-P). The numerical study is performed synergistically with an experimental study using Particle Image Velocimetry (PIV), following an approach similar to the one used in Liberzon et al.~\cite{Liberzon2009}. The synergetic study is important because of two crucial aspects. First, although the FENE-P model proved itself capable to predict the qualitative behavior of turbulent flows of dilute polymer solutions in wall-bounded and quasi-isotropic turbulence, there is little literature available on its applicability in the boundary between turbulent and non-turbulent regions. To the authors best knowledge there is only one previous study of forced shear-less turbulent/non-turbulent interfaces, but this study only addressed a Newtonian case \cite{OberlackGrid}. Second, experiments have shown that in certain conditions, addition of dilute polymers lead to apparently contradictory result: while in most drag reduction applications polymers are found to diminish turbulent fluctuations, in some cases they have been found to amplify them and cause faster propagation rates of turbulent regions~\cite{draad1998,owens2002,Liberzon2009}. 
Therefore, a physically relevant study requires a cross-validation of the qualitative observations on the behavior of the interface of zero-mean-shear turbulence in a dilute polymer solution.

\subsection{Experimental setup}

\begin{figure}[ht]
\begin{centering}
\includegraphics{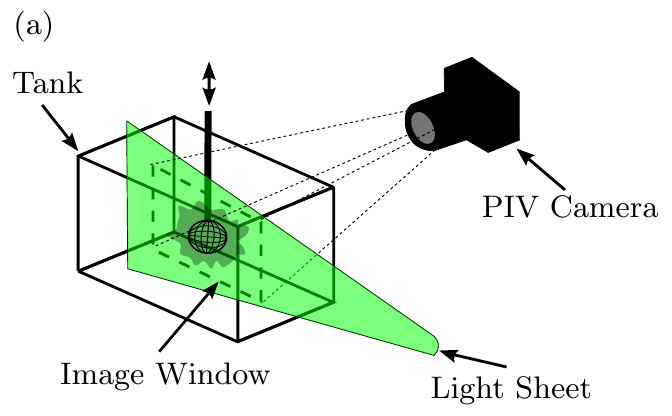}
\includegraphics{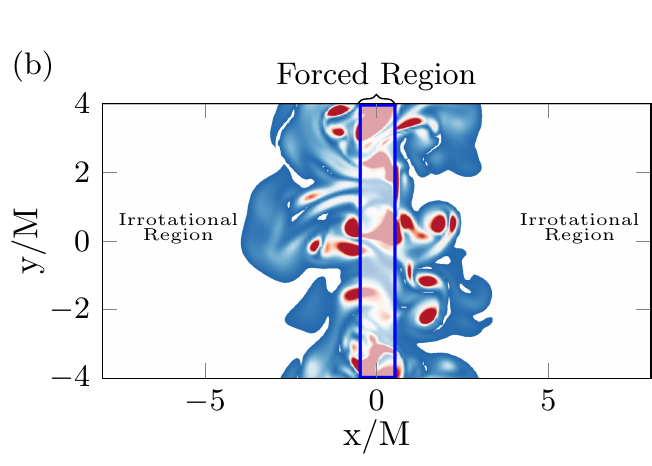}
\caption{(a) Experimental scheme of a turbulent patch, including the PIV laser sheet, camera, spherical oscillating grid agitation. (b) Computational box with a slice of the enstrophy iso-contour for a Newtonian simulation.}
\label{fig:method}
\par\end{centering}

\end{figure}

The physical and numerical experiments are cross-validated in terms of the propagation of quasi-homogeneous turbulent fronts in dilute polymer solutions using the setups shown schematically in Fig.~\ref{fig:method}. In the experiment a localized growing turbulent region is created by a spherical grid. The setup is designed specifically to create a turbulent front while avoiding the wall and shear layer effects that could affect the propagation rates~\cite{Liberzon2009}. The grid has an average mesh size $M = 7$ mm and a vertical stroke of $\pm$ 2 mm is set by a closed-loop controlled linear motor. PIV (synchronized double-head Nd:YAG 120 mJ/pulse laser and 11MP CCD camera, TSI Inc.) measured a flow in a cross-section of an axisymmetric turbulent patch surrounded by the TNTI in water and homogeneous solutions of poly(ethylene oxide) (molecular weight 8$\times 10^{6}$, E-500C, Alkoro GmbH, relaxation time $\tau \approx 7 \times 10^{-3}$ s for the 10 ppm solution). Five independent experiments have been performed for every case. The axisymmetric configuration allows to measure the flow propagating horizontally to the side, reducing the contamination from the vertical fluctuations due to the grid movement. In this way the sidewise propagation of the TNTI is similar to the one modeled in the numerical counterpart. An online supplementary video demonstrates the growth of a turbulent patch in the cross-section as measured by PIV and presented as a $2$-dimensional field of an out-of plane vorticity, $\omega_z$~\cite{youtube}. 
It is worth here anticipating that both the experimental and the numerical set-ups analyze the propagation of turbulence created by a localized 3D forcing. On the other hand, the main difference is that while in the numerical simulation, due to the symmetries of the forcing, the average positions of the TNTI at various times are parallel planes, in the experiment the subsequent average positions considered are spherical zones with increasing radii. While this geometrical difference makes the direct comparison between the properties of the corresponding Newtonian and viscoelastic TNTIs difficult, it does not hinder the possibility to compare the net effect of the introduction of dilute polymers in terms of an enhancement or reduction of the entrainment rate.     

\subsection{Numerical method}

\noindent In the numerical simulation the polymer solution is described in terms of a single relaxation time FENE-P model,  
which reasonably combines feasibility and the capability to reproduce some realistic properties of the solutions. The FENE-P model introduces an extra term on the right-hand side of the Navier Stokes equations
\begin{eqnarray}
g & = & \nu \frac{\partial T^p_{ij}}{\partial x_j}, \nonumber \\ 
T^p_{ij} & = & \frac{\eta_p}{\tau}\left(\frac{L^2_{max}-3}{L^2_{max}-Tr(C_{ij})}C_{ij}-\delta_{ij}\right),
\end{eqnarray}
\noindent where $\eta_p$ is the ratio between the asymptotic zero shear rate viscosity of the solution with polymers and the solvent viscosity, 
 $L^2_{max}$ is the maximum allowed extension of the polymer chain and $C_{ij}$ denotes the polymer conformation tensor. The values chosen for the two parameters
 are reported in table \ref{DNSpara} and they are of the same order of magnitude of the ones used for drag reduction studies \cite{dallas2010,zhang2013,Valente_2014}. The evolution of the conformation tensor reads
\begin{equation}
\frac{\partial C_{ij}}{\partial t}+u_k \frac{\partial C_{ij}}{\partial x_k}=-\frac{1}{\tau}\left(\frac{L^2_{max}-3}{L^2_{max}-Tr(\mathbf{C)}}C_{ij}-\delta_{ij}\right)+\frac{\partial u_{i}}{\partial x_r}C_{rj}+C_{ir}\frac{\partial u_{j}}{\partial x_r}+\chi\frac{\partial^2 C_{ij}}{\partial x_k^2}.
\end{equation}

\noindent The last term on the right hand side is an artificial diffusivity added in order to keep the evolution of the conformation tensor numerically stable. We present the results with  $\chi=0.005$. We have performed extensive tests of the effect of this parameter on the results. For instance increasing the diffusivity to $\chi=0.010$ leads to maximum changes below $2\%$ both for turbulent kinetic energy in the bulk and in the interface propagation.

\noindent The Navier-Stokes and the conformation tensor transport equations are numerically integrated using a pseudo-spectral method de-aliased with the $3/2$ rule and a third-order Runge-Kutta time solver following the implementation used in De Angelis et al.~\cite{DeAngelis2005}. For the simulations $512\times256\times256$ Fourier modes are used for the spatial discretization of a tri-periodic computational box with size $4\pi\times 2\pi\times 2\pi$, while a constant time step $\Delta t=0.002$ has been used for the time discretization. 
The resolution of the numerical simulation is 
 $\eta/\Delta x \approx 1.5$ outside the forced region in the Newtonian case where the Kolmogorov length scale has been estimated as $\eta=(\nu^3/\epsilon_B)^{1/4}$ with $\epsilon_B$ being the dissipation of kinetic energy in the bulk of the flow. The Deborah number of the polymer simulation is defined as $De=\tau/\tau_{\eta}$, with $\tau_{\eta}=\sqrt{\nu/\epsilon_B}$ being the Kolmogorov time scale. 
Further parameters of the simulation together with some of the flow properties at steady state are given in Table~\ref{DNSpara}. The simulation dataset is composed of $10$ independent runs for every case. Bulk statistics of the flow have been sampled at the limit of the region stirred by the body force corresponding to the $y-z$ planes at a distance of $\pm0.6M$ from the center of the domain.
\begin{table}[ht]
\caption{Simulation parameters and average turbulent flow properties in the bulk. $L^2_{\text{max}}$ is the maximum allowed extension of the polymer, $L^2_B$ is $\langle tr(C_{ii}) \rangle$ and $u_B$ are respectively the average of the trace of the conformation tensor and the root-mean-square of the velocity fluctuations $u_i$ in the bulk.}
\begin{tabular}{c c c c c c c c c c c c c}
\hline
\rule{0pt}{2.5ex}    Case & $\nu$  & $\eta_{p}$  & $L^2_{\text{max}}$ &  $\tau$ &M     & $\eta$  &$\tau_{\eta}$ & $M/u_{B}$  & $L^2_B/L^2_{\text{max}}$ & $De$\\ \hline
Newtonian & 0.005  &      &     &       & 0.786 & 0.037 & 0.27      & 1.92  &  & \\ 
Polymer   & 0.005  & 0.1      & 5000 &   2    & 0.786 & 0.041         &   0.35      & 2.02   & 0.20 & 5.6\\ \hline
\end{tabular}
\label{DNSpara}
\end{table} 
Turbulent flow is generated and sustained by the addition of a body force to the right-hand side of the momentum equations. This body force is
tailored to mimic the length scales and the time periodic input produced in an experimental facility with an oscillating grid. For each of the three directions the body force distribution $f_i(x,y,z,t)$ in space and time is determined by the following procedure. First for each component of the body force a random amplitude distribution $A_i(y,z,t)$ in the $y-z$ directions is generated, this is done by assigning random values $\in[-1,1]$ at equispaced nodes with separation $M=2\pi/8$, the amplitude distribution is then obtained in the remaining points of the $y-z$ plane by a bi-cubic interpolation in space intersecting the randomly assigned nodes. A new random distribution is generated periodically with a frequency $1/\mathrm{T}_f$. The passage between two amplitudes distributions in time, $A_i(y,z,n\mathrm{T}_f)$ and $A_i(y,z,(n+1)\mathrm{T}_f)$, with $n\in \mathbb{N}$, is moreover smoothed by interpolating in time the two configurations, which produces a function $\tilde{A}_i(y,z,t)$. The forced region is periodic in the $y-z$ cross section of the domain while it remains confined to a thickness of around $M$ in $x$-direction as it is shown in Fig.~\ref{fig:method}. The final $3$-dimensional time varying distribution of the forcing $f_i(x,y,z,t)$ is given by
\begin{equation}
f_i(x,y,z,t)=\frac{K}{2}\left(1+\tanh\left( \frac{a\Delta}{2}-a|x|\right)\right)\tilde{A}_i(y,z,t),
\label{forcing}
\end{equation}
%
%
\noindent where the parameter $K$ sets the intensity of the body force while $\Delta$ and $a$ determine the thickness of the forced region. In the following, we present the results of the simulation with $\Delta=0.065$, $\mathrm{T}_f=0.1$ and $a=1.5\pi$.
The body force hence imposes an energy injection length scale $M$ which is comparable to the grid mesh size in experiments and a correlation time scale $\mathrm{T}_f$ and 
the virtually infinite turbulent front propagates in the $x$-direction as shown in Fig.~\ref{fig:method}b. 

\subsection{Choice of comparable runs}

\begin{figure}[ht]
\begin{centering}
\includegraphics{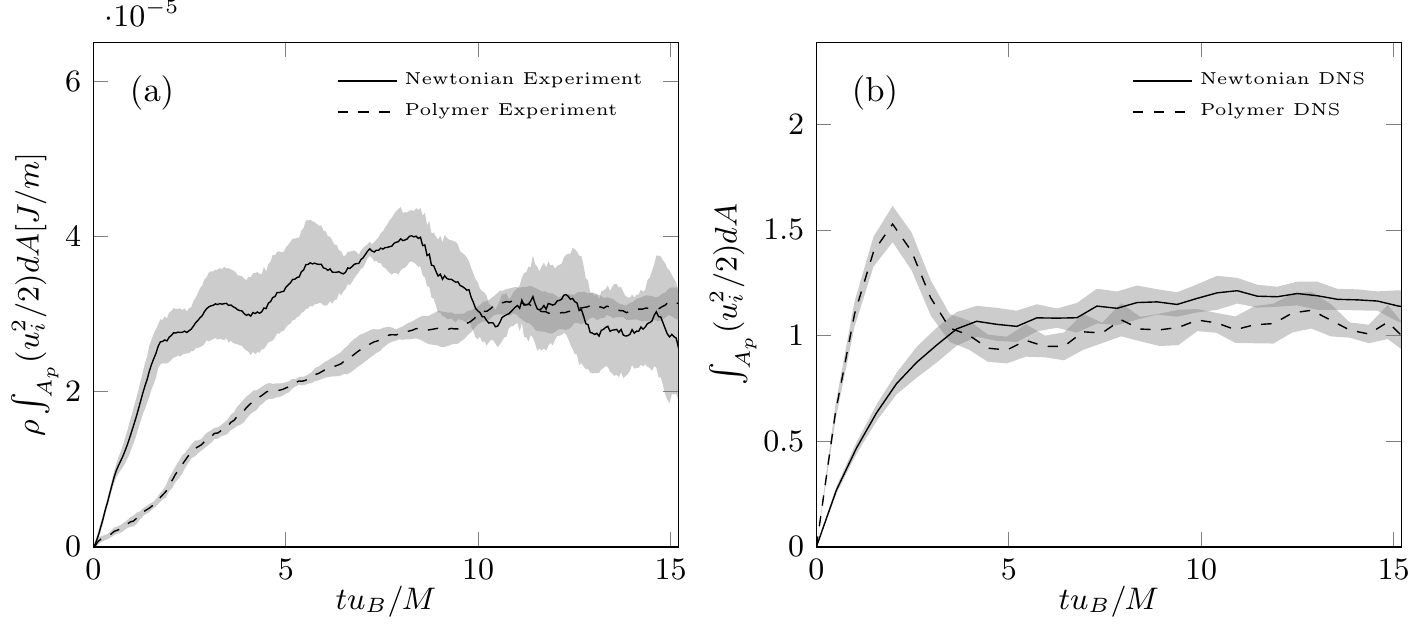}
\caption{(a) Kinetic energy integrated over the cross section of the turbulent region in experiments. Experiments correspond to 6.9 Hz in water and 10.5 Hz for 10 ppm polymer solution, respectively. (b) Kinetic energy integrated over the cross section of the turbulent region in DNS. Grey areas correspond to $\pm 1.96$ standard errors of the mean estimated from the variance of the ensemble. }
\label{fig:energy}
\par\end{centering}
\end{figure}

Whenever two flow states are to be compared one needs to make a choice on how to set up the comparison~\cite{hasegawa2014}. In the present work we aim at investigating the effect of polymers on the turbulent entrainment process under the condition that this process  ``receives'' the same energy supply. The energy supply to the interface is measured through the integrated turbulent kinetic energy contained in the turbulent patch. Since it is not possible to determine the turbulent kinetic energy content of a turbulent patch a priori, multiple runs with and without polymers have been done in experiment and numerics. Afterwards, runs in which the turbulent kinetic energy within the turbulent regions of water and dilute polymer solution match in amplitude are selected for comparison. A set of corresponding examples is shown in  Fig.~\ref{fig:energy}a and \ref{fig:energy}b.

In the PIV experiment runs with different oscillation frequencies were carried out. Due to the axisymmetric shape of the turbulent patch, it is actually possible to measure a thin slice of the three-dimensional patch only. Hence the two components of the velocity $u_{1,2}$ are squared and integrated over the cross-sectional area of the patch, $A_p$, thus $\frac{1}{2} \rho \int_{A_p} u_i^2 dA$ provides the total kinetic energy of the points belonging to the turbulent patch. The total kinetic energy in the slice, shown in Fig.~\ref{fig:energy}a, grows in time (in some cases it has an overshoot and decrease due to strong vortices leaving the observation volume~\cite{youtube}) and eventually reaches a quasi-stationary stage.

In the DNS the integrated kinetic energy in the patch also reaches a quasi-steady state. Different runs are realized by adjusting the forcing amplitude $K$ in equation (\ref{forcing}) and - just like in the experiment - two cases with similar steady-state-levels of kinetic energy are selected.

During the initial transient in the DNS for the polymer case a steep growth of energy is observable in Fig.~\ref{fig:energy}b which is caused by a stronger forcing action (required to achieve the same level of energy for the steady state). After peaking at around 2 eddy turnover times the energy drops to the quasi-stationary value. This has been assumed here to be the consequence of the fact that at the grid walls the sharp velocity gradients necessary to stretch the polymers are generated immediately in the boundary layer forming over the grid walls. In the body force model such gradients are produced only throughout the evolution of the turbulent cascade which is not an instantaneous process.
Similar features can be observed in statistics of other transient viscoelastic simulations \cite{dubief2001, min2003,perlekar2010}. The results presented in this paper are based on the quasi-stationary stage only and time-averaging is performed starting from $tu_{B}/M=7$ where $u_B$ is the root mean square of the velocity fluctuations in the bulk.

The Reynolds number is defined as $Re=u_{B} M/\nu$ (using the solvent viscosity). In the following, only the data for the case with $Re \approx 67$ will be shown for both experiments and simulations. In the simulations this has been estimated to correspond to a value of $Re_\lambda=u_{B} \lambda/\nu$ of about $50$ for the Newtonian case, where the Taylor microscale has been computed as $\lambda=(15 \nu u_{B}^2/\epsilon_B)^{1/2}$.
%
In the experiment $Re \approx 67$ was obtained with forcing frequencies of $6.9$ Hz for the Newtonian case and $10.5$ Hz for the polymer case at a 10 ppm concentration. In the simulation $K$ is set equals $3.8$ for all the Newtonian runs and $5.85$ for all the runs with polymer. 

The average position of the interface $X_I(t)$ is computed as a spatial average over all the positions of the points $x_I = x_I (y,z)$ of the interface. The interface is detected by finding the outermost point where enstrophy, $\Omega = \omega_i \omega_i / 2$ ($\omega_i$ denotes vorticity), equals a given threshold $\Omega_{th}$. The value of the threshold used for the interface detection is $2\%$ of the maximum of the mean enstrophy at a given time, i.e. $\Omega_{th} = 0.02 \Omega_b$. In such a way in the simulations an isosurface of enstrophy in a $3$-dimensional space is identified and in the following part of the paper, the statistics referring to the interface are obtained by averaging over the points of this isosurface.
\begin{figure*}[t]
\begin{centering}
\includegraphics{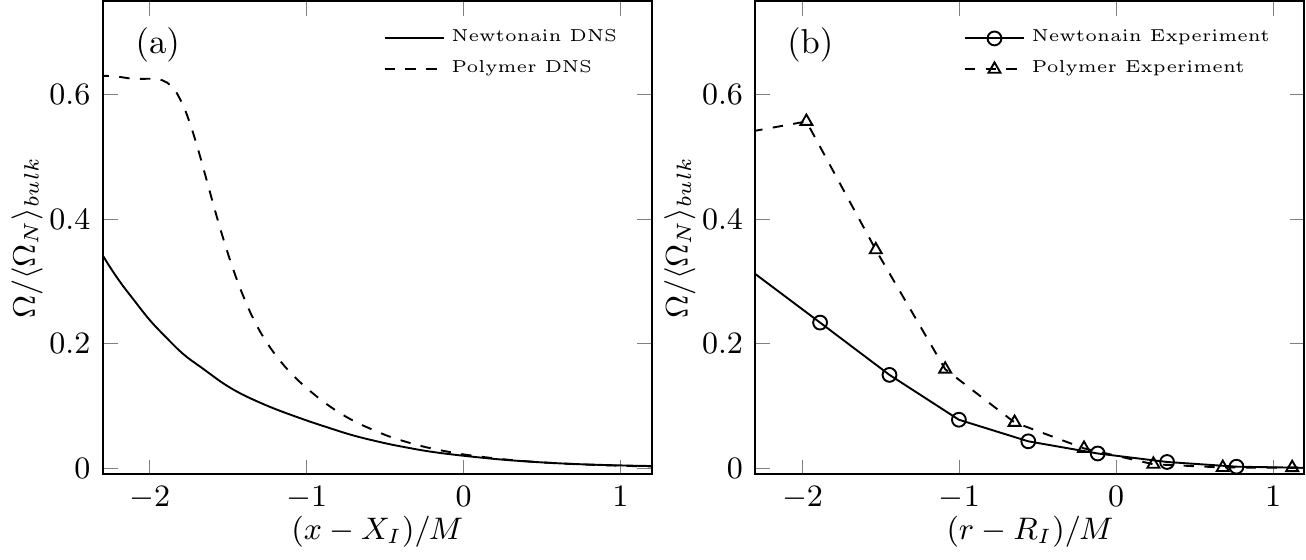}
\caption{Mean enstrophy as a function of the distance from the average interface position normalized by the average enstrophy of the Newtonian case in the bulk. (a) DNS, (b) Experiments; the values are normalised by the average enstrophy for the Newtonian case in the bulk.}
\label{fig:enstrophy}
\par\end{centering}
\end{figure*}

\begin{figure}[!ht]
\begin{centering}
\includegraphics{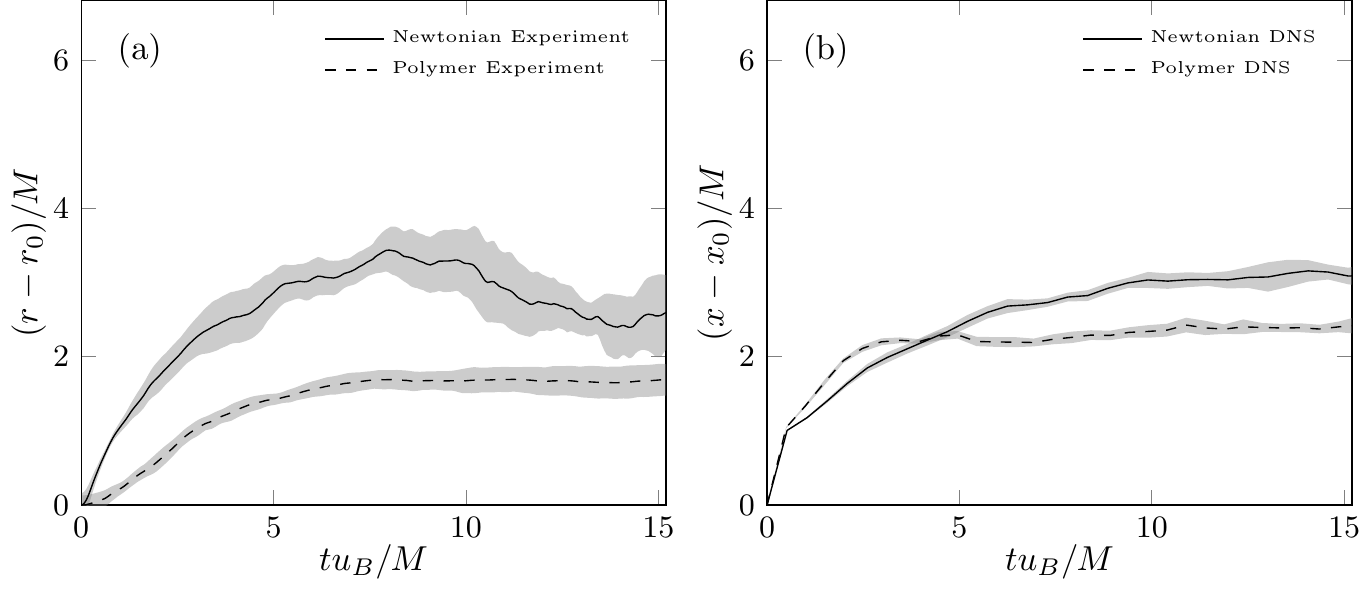}
\caption{(a) Average interface position with time in experiment. Experiments correspond to 6.9 Hz in water and 10.5 Hz for 10 ppm polymer solution, respectively. (b) Interface position with time in DNS. Grey areas correspond to $\pm 1.96$ standard errors.}
\label{fig:size}
\par\end{centering}
\end{figure}

\noindent In both the simulations and experiments the region close to the interface experience different enstrophy distribution with polymers compared to the Newtonian case as can be seen from Fig.~\ref{fig:enstrophy} where the values normalized with the average enstrophy in the Newtonian bulk show that vorticity close to the TNTI is larger in the polymer case. Nevertheless the propagation appears to be less effective in the flow with polymers. Indeed the equivalent size of the turbulent region is shown in Fig.~\ref{fig:size}a and \ref{fig:size}b, when the patch reaches a stable size, for the polymer case the values are around $1\div2$ mesh sizes smaller than their Newtonian equivalent. In the experiment, the results are presented in terms of an effective radius $r$, estimated using the accurately measured area of the turbulent region and presented as a segment of an axisymmetric region.

\noindent It is noteworthy that in previous studies under an oscillating grid spanning the full width of the tank, the growth rate of a turbulent region in polymer solution was in some cases faster and in some cases slower than the Newtonian case~\cite{Liberzon2009}. 
 In the present study we find both in experiments and in DNS that a finite size patch is reached with the localized forcing. For the same level of turbulent kinetic energy contained in the patch this equilibrium patch size is smaller for the dilute polymer solution than for the Newtonian case. This observation indicates that the turbulent flow in polymer solutions entrains less fluid across the TNTI when the same level of turbulent kinetic energy is available within the patch (as pointed out before, these equivalent levels of kinetic energy are posteriori selections from multiple tests with different stirring intensity).  

The interfaces produced in the polymer cases also visually appear to be less convoluted and smoother, both in simulations and experiment. Unfortunately the geometric differences do not allow for a quantitative comparison of the experiment and simulations regarding the smoothness of the interface. 
 
\section{Results}\label{sec:results}

After the selection of the two cases for Newtonian and polymer solution cases, we can use DNS results to study in depth the mechanisms for the observed difference in the entrainment. Since polymers interact with the flow via its velocity gradients, it is natural to focus on the velocity gradient tensor in its symmetric and antisymmetric parts, i.e. the rate-of-strain (hereinafter called strain for the sake of brevity) and vorticity. The strain $s_{ij}$  is known to be depressed in drag reducing fluids in general, and in dilute polymer solutions in particular,\cite{Liberzon2005} but has not been studied at TNTI before.

\begin{figure}[ht]
\includegraphics{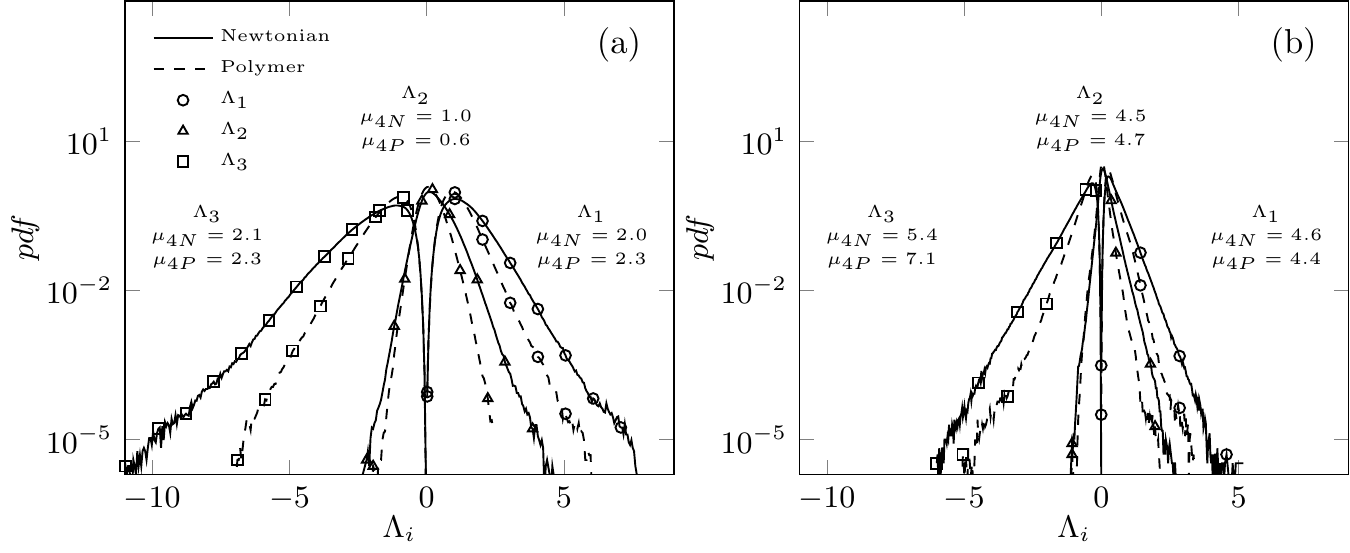}
\caption{Probability density function of the eigenvalues of the rate of strain tensor in the DNS. Left: in the bulk of the flow; right: over turbulent/non-turbulent the interface. Values of the kurtosis $\mu_4$ are given for each curve on the plot.} \label{fig:straineigval}
\end{figure}

A representation of the strain field eigenframe allows a deeper look into its internal dynamics. We start with the probability density function (\textit{pdf}) of the eigenvalues of the strain rate ($\Lambda_1 \geq \Lambda_2 > \Lambda_3$, where $\Lambda_3 < 0$ and $\Lambda_1^2+\Lambda_2^2+\Lambda_3^2 = s_{ij}s_{ij}$)~\cite{Tsinober2009}. We present these \emph{pdf} in the bulk and at the interface as depicted in Fig.~\ref{fig:straineigval} (a-b). As it was previously observed experimentally~\cite{Liberzon2005} the \emph{pdfs} are qualitatively similar in both Newtonian and viscoelastic turbulent flows. The difference are only in the reduced tails of the \textit{pdf} for all the eigenvalues of the polymer case. The average values of $\Lambda_i$ are shown in Table \ref{table_eig} reflecting the reduction in the eigenvalues intensity in both the bulk and at the interface when the polymers are introduced. 

At the interface, the frequency of strong strain events is reduced compared to the bulk and the kurtosis $\mu_4$ of the three distributions is increased as shown in Fig.~\ref{fig:straineigval}. The change in the PDF of 
 $\Lambda_2$ for the polymer flow compared to the Newtonian one is particularily interesting since the polymers appear to reduce the positive events more than the negative ones (reduced skewness of  $\Lambda_2$). This change in the PDF, which is observable in the bulk and at the interface, implies a reduction of the positive contributions to the enstrophy production, (compared to the Newtonian case) and hence on average a minor value of the latter, as can be shown feeding the values of Table \ref{table_eig} in the relationship with the the strain rate eigenvalues  $\Lambda_1\Lambda_2\Lambda_3=-1/4 \,\omega_i\omega_js_{ij}$ \cite{Tsinober2009}.

\begin{table}[h]
\centering
\begin{tabular}{c cc  cc}
\hline
\rule{0pt}{2.5ex} & \multicolumn{2}{c}{Bulk} & \multicolumn{2}{c}{Interface} \\
\multicolumn{1}{l}{} & \multicolumn{1}{l}{Newtonian} & \multicolumn{1}{l}{Polymer} & \multicolumn{1}{l}{Newtonian} & \multicolumn{1}{l}{Polymer}\\ \hline
\rule{0pt}{2.5ex} $  \langle \Lambda_1 \rangle   $ & 1.34 & 1.12 & 0.49 & 0.39 \\
\rule{0pt}{2.5ex} $  \langle \Lambda_2 \rangle   $ & 0.29 & 0.15 & 0.14 & 0.07 \\
\rule{0pt}{2.5ex} $  \langle \Lambda_3 \rangle   $ & -1.70 & -1.28 & -0.64 & -0.47 \\ \hline
\end{tabular}
\caption{Average values of the strain rate eigenvalues  from the DNS.}
\label{table_eig}
\end{table}
\begin{figure}[]
\centering
\includegraphics{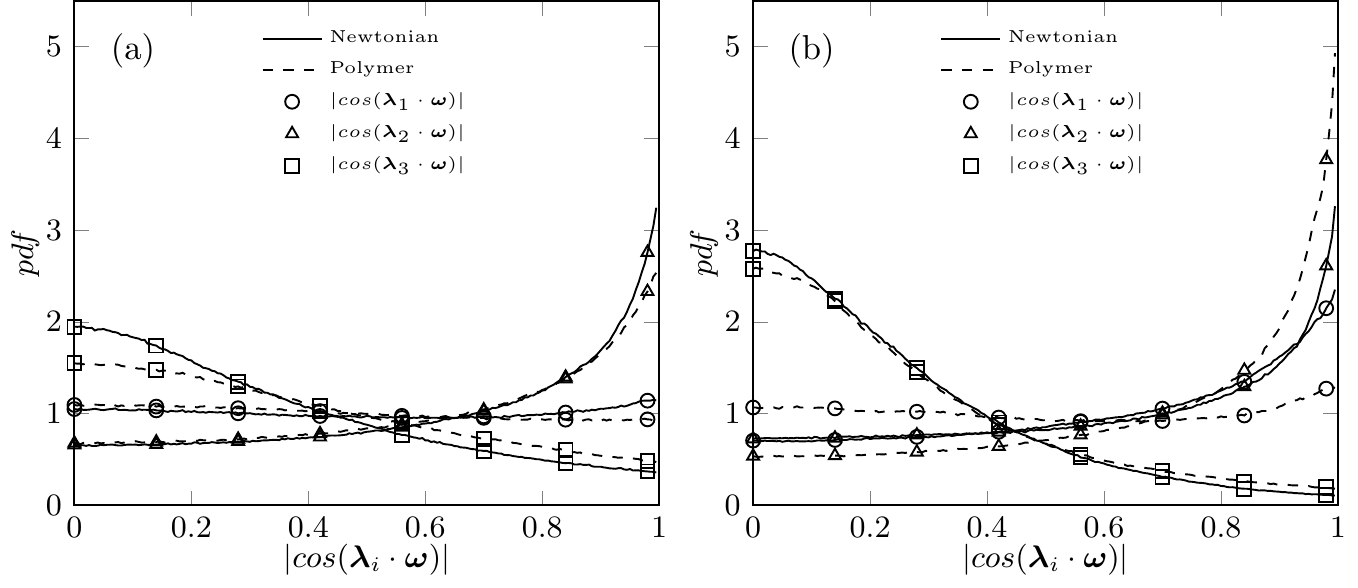}
\caption{Probability density function of the cosine of the angle between the vorticity vector and the strain rate eigenvectors in the DNS. A value of $1$ represent perfect alignment between ${\lambda}_i$ and  ${\omega}$ while for values of $0$ the two vectors are orthogonal. (a) in a turbulent bulk, (b) at the interface. }
\label{fig:teta_strain_vort}
\end{figure}

We continue with the analysis of the alignment between vorticity and strain of Fig.~\ref{fig:teta_strain_vort}. The alignment is known to be strongly linked to the dynamics of both strain and enstrophy production and destruction~\cite{Tsinober2009,luthi2005}. We know that in genuinely turbulent flows $\omega$ is predominantly aligned with the eigenvector $\boldsymbol{\lambda}_2$ (corresponding to $\Lambda_2$) and that enstrophy production depends, as well as on the rate of dissipation, on the geometrical alignments, i.e. $\omega_i \omega_j s_{ij} = \omega_i^2 \Lambda_i \cos^2(\boldsymbol{\omega},\boldsymbol{\lambda}_i)$\citep{Tsinober2009}. In Fig.~\ref{fig:teta_strain_vort} (a) we see that in the bulk of the flow there are only minor differences in alignments between the Newtonian case and the polymers. For both types of fluids the vorticity remains aligned with $\boldsymbol{\lambda}_2$. However, at the interface we can observe two interesting phenomena linked to the changes observed in the $pdf$ of strain eigenvalues. First, for the polymeric fluid the alignment with $\boldsymbol{\lambda}_1$ appears much lower than the one of the Newtonian case at the interface. As  L\"uthi et al.~\cite{luthi2005} have shown, the strongest contribution to $\omega_i \omega_j s_{ij}$ comes from $\omega^2\Lambda_1\cos^2( \boldsymbol{\omega} \boldsymbol{\lambda}_1$), hence from $\boldsymbol{\lambda}_1$-alignment. A reduction observed in the polymer case points to a strongly reduced stretching of vorticity. Second, for the polymer case, we observe an increased alignment with $\boldsymbol{\lambda}_2$. A stronger $\boldsymbol{\lambda}_2$-alignment leads to a small, yet on average positive, contribution to the enstrophy production.
The fact that polymers lead to a reduced alignment with the
strong stretching eigenvector at the interface while increasing the alignment with the intermediate
eigenvector suggests that they alter the entrainment mechanism.

\begin{figure}[]
\centering
\includegraphics{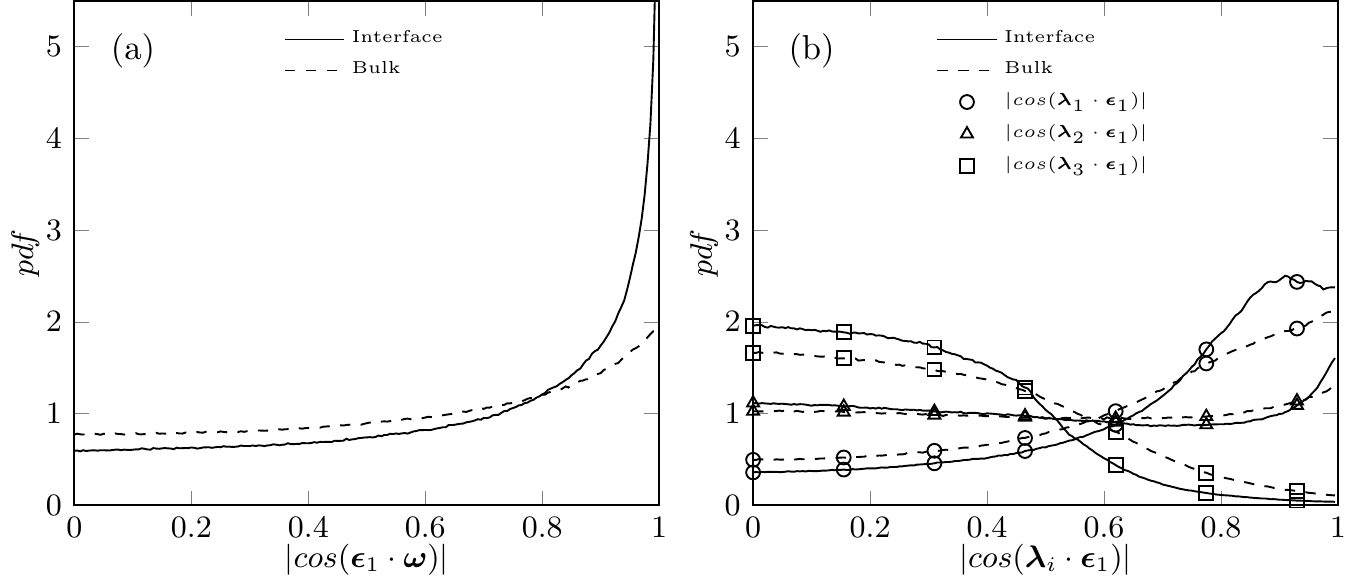}
\caption{Probability density function of the cosine of the angle between (a) $\epsilon_1$ and the vorticity, (b) $\epsilon_1$ and the strain rate eigenvectors. Both refers to the DNS for the polymer case.}
\label{fig:teta_strain_poly}
\end{figure}

It is possible to link the reduction of strain eigenvalues and the change in alignment with vorticity to the changes in alignment between the polymer conformation tensor with both vorticity and strain eigenframe. Indeed in Fig.~\ref{fig:teta_strain_poly} (a) it is possible to observe that the largest polymer eigenvector $\boldsymbol{\epsilon}_1$ representing the polymer orientation is preferably aligned in the bulk with the vorticity vector confirming what was previously observed in homogeneous isotropic turbulence \cite{Vincenzi2015,watanabe2010}. Moreover it can be seen that at the interface polymers align stronger with the vorticity indicating that the polymers tend to be oriented parallel to the interface. In this scenario, the polymer stress is expected to affect vorticity indirectly via changes to the strain field. This can be explained by observing that in the FENE-P model the action of the flow field on the polymers is accounted by the terms $\frac{\partial u_{i}}{\partial x_r}C_{rj}+C_{ir}\frac{\partial u_{j}}{\partial x_r}$ which are directly dependent on the rate-of-strain only, making these equivalent to $s_{ir}C_{rj}+C_{ir}s_{rj}$ \cite{Tsinober2009}. The polymer stress depends hence on the local equilibrium between the contribution from the strain field, which tends to stretch the polymers, and the restoring elastic force. When the polymer is aligned with vorticity the stress is expected to produce a reduced stretching effect on vorticity and thus on its production through the $\omega_i \omega_j s_{ij}$ term. 

\begin{figure}[ht]
\includegraphics{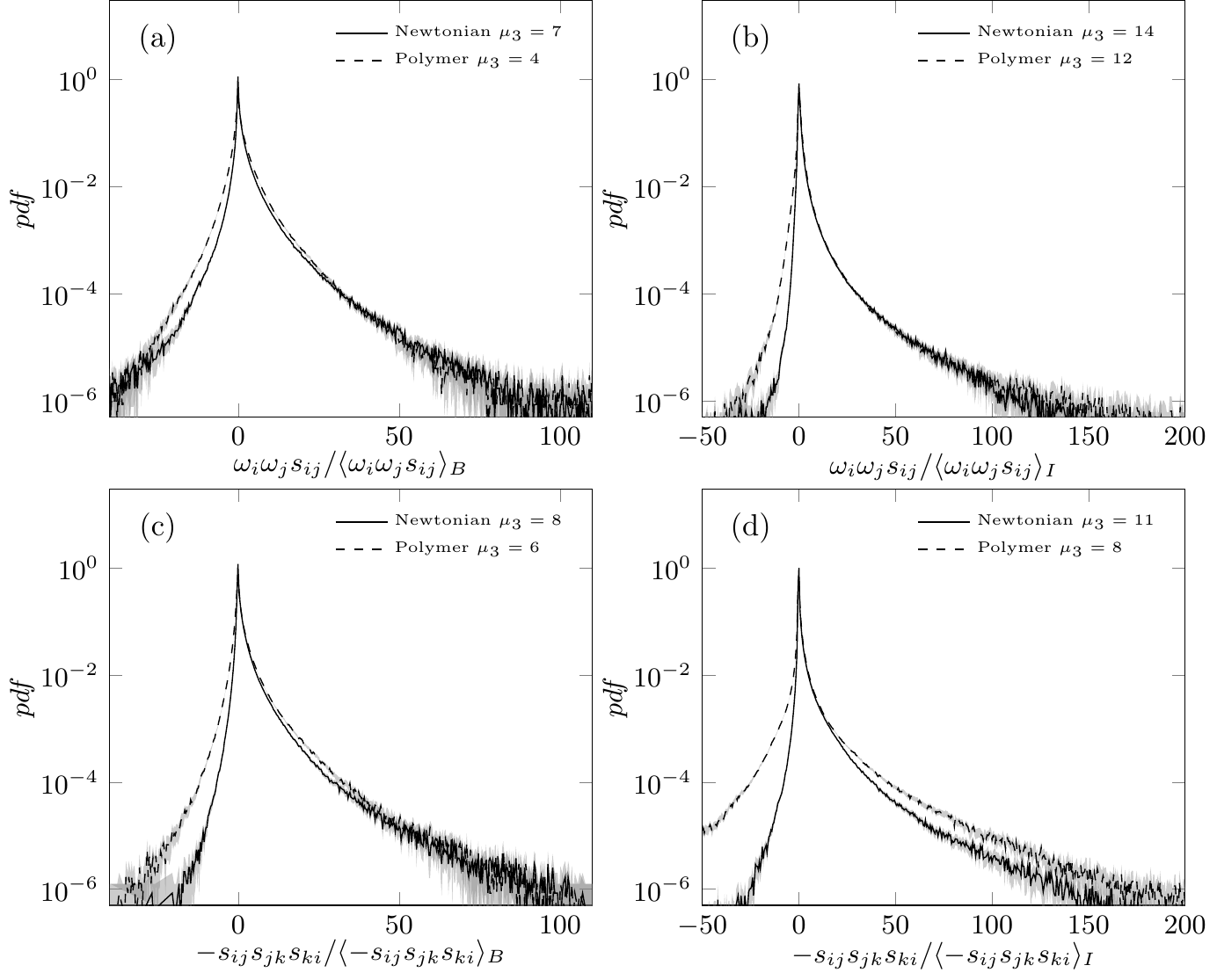}
\caption{Above: DNS results for the probability density function of the enstrophy production $\omega_i\omega_js_{ij}$ normalized by its average value at the given position in (a) the bulk of the flow, (b) over the turbulent/non-turbulent interface. Below: DNS results for the probability density function of the strain rate production $-s_{ij}s_{jk}s_{ki}$ normalized by its average value at the given position in (c) the bulk of the flow, (d) over the turbulent/non-turbulent interface. Grey areas delimits $\pm 1.96$ standard errors of the mean estimated using a Poisson approximation.} 
\label{fig:production}
\end{figure}

The modifications in the alignments between vorticity and the strain eigenframe observed at the interface in Fig.~\ref{fig:teta_strain_vort} (b) are thus primarily determined by the way polymers alter the strain eigenframe. Fig.~\ref{fig:teta_strain_poly} (b) present the \textit{pdf} of the relative orientation between the largest polymer eigenvector $\boldsymbol{\epsilon}_1$ and the strain eigenvectors $\boldsymbol{\lambda}_{1,2,3}$ respectively for the bulk of the flow and for the points on the TNTI. The polymers are preferentially aligned with $\boldsymbol{\lambda}_1$ ($\Lambda_1 > 0$, stretching eigenvector) in the turbulent bulk again in agreement with the results from homogeneous isotropic turbulence with polymers~\cite{Valente_2014,Vincenzi2015,watanabe2010}. At the interface the alignments with the eigenframe changes: there is an increase in $\boldsymbol{\lambda}_1$ alignment and decrease of alignment with the compressing eigenvector $\boldsymbol{\lambda}_3$. Although the changes in $pdf$ are not striking, the subtle alterations produced by the polymer into the strain eigenframe and on its orientation are significantly reflected in the balance of enstrophy and strain production, as shown in Fig.~\ref{fig:production}. 

Fig.~\ref{fig:production} shows the \textit{pdf} of the enstrophy production $\omega_i \omega_j s_{ij}$ and strain production $s_{ij} s_{jk} s_{ki}$ terms, comparing the turbulent bulk results with the distributions at the interface. The polymer elongation (not shown for the sake of brevity) is obviously larger in the turbulent bulk where the velocity gradients are more intense. However, even relatively weaker polymer elongation near the interface has a particularly significant effect on the small-scale processes there. This is evident in the redistribution of the enstrophy-strain dynamics that at the interface are responsible for the amplification of non-linearities in the fluid entrained via viscous diffusion. In the bulk of the flow the normalized enstrophy production show no sensible variation in the balance between straining (positive) events and compressive (negative) ones. Closer to the interface it is possible to observe that polymers locally affect the production of enstrophy by shifting the balance between compressive and stretching events towards the compressive ones.

As expected, the strain rate production is more evidently affected by polymers and we observe a larger imbalance towards negative events, in the bulk and even more intensely at the interface. As previously observed this is linked to the changes in alignment between vorticity and strain in the polymer flow. The strong $\boldsymbol{\lambda}_2$ alignment seems somewhat to compensate the reduced contribution from $\boldsymbol{\lambda}_1$ alignment in the positive side of the $pdf$ of the enstrophy production. A stronger $\boldsymbol{\lambda}_2$ alignment also implies a stronger contribution from its negative events, at the same time positive events linked to  $\boldsymbol{\lambda}_2$ alignment are weaker globally leading to an increased weight of the negative side of the $pdf$ of the enstrophy production.

The process of turbulent regeneration is directly dependent on the balance of both $\omega_i \omega_j s_{ij}$ and $s_{ij} s_{jk} s_{ki}$ as it can be seen from the equation for the rate of strain (forcing and inviscid production through pressure Hessian are omitted for the brevity)~\cite{Tsinober2009}: 
\begin{equation}\label{eq:strain_production}
\frac{1}{2} \frac{Ds^2}{Dt} = -s_{ij}s_{jk}s_{ki} - \frac{1}{4} \omega_i \omega_j s_{ij} +\nu s_{ij} \nabla^2 s_{ij}  \ldots 
\end{equation} 
The change in balance is weak in the bulk of the flow, as can be observed from the skewness values ($\mu_3$ shown in Fig.~\ref{fig:production}). However, it becomes significant at the interface where strain production dominates over enstrophy production\cite{dasilva2008} and where vorticity, predominantly oriented parallel to the interface\cite{gampert2014}, is strongly anisotropic. In such a situation the space of possible interaction between strain, polymers and vorticity is effectively limited resulting in a more coherent alignment which is a possible reason for the amplified effect of polymers.
As it was previously inferred~\cite{tsinober90b,Tsinober2009} and as also observed here, polymers affect the interface through (mis-)alignments of the strain eigenvectors and vorticity vectors and this leads to reduced non-linearity and decreased production of strain. This reduced production of strain, in turn, affects vorticity production, and coupled with a strongly modified alignment of strain and vorticity, reduces the capability of the turbulent flow to entrain the non-turbulent fluid. 

\section{Conclusions}\label{sec:conclusions}

We studied the properties of a shearless turbulent/non-turbulent interface in a dilute polymer solution focusing on properties of orientation between strain, vorticity and polymers. While such informations are easily accessible in numerical simulations with polymers, the underlying assumptions contained in the polymers model are known to not always lead to physical results \cite{ghosh2001}. Thus the applicability of the FENE-P model in the framework of shearless turbulent/non-turbulent interfaces has been carefully cross-validated with the experimental results.
In both experiments and simulations polymers appear to reduce the maximum size reached by the turbulent patch and produce interfaces with smoother features  even when the turbulent kinetic energy available in the patch is comparable.

While polymers affect the turbulence everywhere in the flow, strain-vorticity orientation statistics seem to be little affected in the bulk of the flow and follow the patterns previously observed in a number of different flows with and without polymers \cite{Vincenzi2015,Valente_2014,elsinga2010}. The universal character of the orientation between vorticity and strain eigenframe can also suggest that dynamics of self-amplification of vorticity fluctuations are not strongly affected by the polymers in the bulk and most of the alteration in turbulence comes from the strain field with which polymers directly interact.
Strain-vorticity alignments, and as consequence enstrophy production, appear to be more strongly affected at the turbulent/non-turbulent interface. There the polymers show how they can affect a turbulent flow via minor but significant alterations in the dynamics of production of strain and enstrophy. These alterations manifest the importance of the mechanisms of creation and self-sustaining reproduction of turbulence in the process of turbulent entrainment in addition to viscous diffusion.

The turbulent/non-turbulent interface is characterized by strong instantaneous shear, anisotropy and organized vorticity. The reorganization of turbulence leads to a more coherent alignment of polymers with vorticity (and thus with the interface), showing different interaction patterns with the strain and vorticity and an increased weight of vortex compression compared to the Newtonian case. In this sense the results from the present work are consistent with the observations made for turbulent channel flows where polymers in the near wall region preferentially align with the wall \cite{bagheri2012} and interact with the vortex streaks\cite{Dubief_2004}, while in the middle of the channel they settle to orientations similar to the ones found in isotropic turbulence\cite{bagheri2012}.

\begin{acknowledgments}
The authors gratefully acknowledge the support through German-Israeli Foundation for scientific research and development. GC and BF acknowledge additional support through DFG project FR 2823/5-1.
The authors also acknowledge support by the state of Baden-W\"urttemberg through bwHPC.
\end{acknowledgments}
%


%

\end{document}